\documentclass[aps,prd,showpacs,eqsecnum,amsmath,amssymb,nofootinbib,onecolumn]{revtex4}
\usepackage[utf8]{inputenc}
\usepackage{graphicx}
\usepackage{subfigure}
\usepackage{array}
\usepackage{multirow}
\usepackage{amsmath}
\usepackage{amssymb}

\newcommand{\bO}{{\boldsymbol \Omega}}

\newcommand{\be}{{\boldsymbol \epsilon}}
\newcommand{\sinc}{\text{sinc}}

\usepackage{hyperref}
\usepackage{cleveref}

\begin{document}

\title{Laser interferometer response to scalar massive gravitational waves}
\author{Arkadiusz B{\l}aut}
\affiliation{Institute of Theoretical Physics, University of Wroc\l aw, Wroc\l aw, Poland}

\begin{abstract}
We analyze the response of the gravitational wave detector to a scalar massive plane gravitational wave.
We give the compact form of the response and discuss its angular and frequency characteristics.
The derivations is carried out in the conformal and the synchronous gauges and the equivalence of the two approaches is shown.
In the case of the massive Brans--Dicke theory we solve the linearized vacuum field equations in the two gauges as well.
\end{abstract}


\maketitle

\section{Introduction}

Alternatives to the General Theory of Relativity (GR) were present already in the early times of its birth and
subsequent periods of its development but a particular attention they have been drawing in the last two decades.
The early attempts (see \cite{WillBook} and references therein) aimed at founding a theoretical framework for the
ongoing weak--field, low--energy experiments in the Earth and the Solar System gravitational environments,
strong--field astrophysical observations like those of pulsars and compact binary systems and then future gravitational wave detection experiments \cite{Eardley73} exploring dynamical, strong--field regime of gravity. The principal theoretical
motivation for studying extensions of GR have been the unification of gravity with the rest of the Standard Model (SM)
interactions \cite{Goenner04} and the formulation of the consistent quantum theory of gravity -- both issues still await
fully satisfactory solutions. Nevertheless the studies of high energy unification models (string theory, supergravity,
effective theories emerging from quantum gravity, brane-world models, extra dimensional models, theories with noncommutative geometries, theories based on deformed Lorentz symmetries) predict deviations from GR at some high energy level and in a lower--energy limit lead to a number of competing effective theories. But can we expect modifications of GR at low energy scale as well?
This cannot be excluded: although GR as far has passed safely most of the local weak--field and strong-field astrophysical
tests it is well known that it needs an additional hypothetical contribution known as a Dark Matter to explain the galactic and cluster scale observations of the rotation profiles \cite{Milgrom83},\cite{TeVeS}. Recently the new motivations has come from cosmology: observations of the Supernovae Type Ia \cite{Riess2004} and Cosmic Microwave Background radiation anisotropies \cite{Planck2014} revealed the accelerated expansion of the Universe favoring the $\Lambda$CDM model with another "dark" component dubbed as Dark Energy. It is also well known that modified gravity can play a role in the early--time cosmology. The first inflationary scenario was proposed in the Starobinsky $f(R)=R+R^{2}/6M^{2}$ model \cite{Staro80}; improvement of the chaotic inflation through the non-minimal coupling $\zeta\Phi^2\,R$ of gravity and matter was proposed in \cite{FaUn88}; interesting example of the unified picture of the gravity and SM particle physics was given in \cite{Shapo08}, where it was argued that the SM Higgs field $H$ strongly non-minimally coupled to gravity, $\zeta HH^{\dagger}R$, $\zeta\gg 1$, can give rise to inflation. (For a review of modified gravity in cosmology see e.g. \cite{CFPS2012}).
Furthermore one expects that working ground--based gravitational wave detectors such as the Advanced Laser
Interferometer Gravitational Observatory (Ad.LIGO) \cite{AdLIGO} or planned next generations Einstein Telescope (ET) \cite{ET} and space--based evolved Laser Interferometer Space Antenna (eLISA) \cite{eLISA} will give an unique opportunity to test GR and possibly to find some interesting observational challenges pointing to a new gravitational physics. In this respect it is quite important to know the signals at the detector to discern between competing theories.

In this paper we analyze the response of the gravitational wave interferometer to a gravitational wave signals that arise
in scalar--tensor (ST) theories (see e.g. \cite{FM2003}). In those theories gravitational interactions are mediated by
nonminimally coupled massless helicity $2$ field and a scalar field. In the most studied example, the Brans--Dicke (BD) theory \cite{BD61}, the scalar field was massless but recently the massive scalar fields have been investigated for a potential role they can play in the early and late time cosmology and in astrophysics \cite{YPC12}, \cite{ABWZ12} \cite{Jac99}. It was also recognized that Brans--Dicke theory with the BD parameter $\omega_{BD}$ equals to zero contains as a a subclass the so called extended theories of gravity with gravitational action defined by some function $f(R)$ of the scalar curvature \cite{SF2010}, \cite{FT2010}. In both those theories matter fields are coupled minimally to the tensor field $g_{\mu\nu}$. Dynamics of the metric field however differs from the dynamics in GR due to the nontrivial interaction of the metric with the scalar field in the ST theories and due to the modified field equations in the case of extended theories. For the gravitational waves this shows up as an additional, in general massive, spin $0$ mode of the wave that potentially can be detected by interferometers.
Gravitational waves were also studied in other generalized theories. Recently an exact plane wave solution was found in \cite{Bab12} for a class of Horndeski theory \cite{Horn74} which can be considered as a generalization of the scalar-tensor Brans-Dicke theory and in \cite{Mohseni11} in the case of the nonlinear massive dRGT gravity model \cite{dRGT11}.

The nonrelativistic sector of gravitational interactions in these theories is also modified. For example if the scalar field in massive BD theory satisfies the Klein--Gordon equation with the mass parameter $m$ one expects that the effective Newtonian gravitational potential in the near zone of a source will be modified by the Yukawa--type corrections, $\sim \exp{(-m r)}$. This corrections would manifest themselves as a deviation from the Keppler's third law and can be investigated by observing the dynamics of the planets of the Solar System \cite{BGS2011}. The uncertainty of those measurements can be interpreted as providing the upper bounds on the mass parameter. The strongest upper bound, $m<4.4\times 10^{-22}$eV, comes from observations of Mars \cite{Will98}. But whatever the Solar System bounds might be one must take notice that in theories predicting massive scalar waves the mass parameter may have a dynamical origin. For example in massive ST theories it is defined by the local minimum of some potential $V(\phi)$ which determines the dynamics of $\phi$. However one can also consider potentials having a number of local extrema which, depending on the external conditions and directly on the value of the scalar field, could lead to different dynamically generated masses. This would not be an unusual scenario and in fact it is analogous to the SM Higgs mechanism for spontaneous mass generation. The desired nonperturbative effect in ST theory would be the {\it scalarization} phenomenon in neutron stars where nontrivial configurations of a large scalar field can appear \cite{Damour93}. Another example comes from Einstein--Aether theory which predicts gravitational waves of different polarizations and different propagation speeds although all modes are massless
i.e. their frequencies are proportional to wave vectors \cite{Jac04}.
This examples illustrate that the relativistic, strong field domain may have quite distinctive features than those predicted or extrapolated from nonrelativistic and low--energy range and shows that the gravitational waves may be good probes in exploring this regime.

The response of the gravitational wave detectors to the scalar mode was investigated already in \cite{Shibata94}, \cite{MagNic00}, \cite{Nakao01}, \cite{Corda07} and detection capability together with astrophysical and cosmological application were presented in \cite{Nakao01}, \cite{CCL08}, \cite{BCLF09}, \cite{BoGa05}. Here we continue these efforts and further analyze the detector response. We present close form of the detector response for massive scalar perturbations that straightforwardly reveals the angular and frequency characteristics of the antenna in the whole frequency domain.
The detector response is obtained by analyzing the motion of the the emitter, detector and laser light in the conformal gauge and synchronous gauge. Working in the synchronous gauge in which the free motion of test particles (thus e.g. emitters, beam--splitters of a freely falling gravitational wave detectors) can be easily computed is particularly convenient for space--based interferometers where the high frequency domain of the detector response usually plays an important role. Furthermore we show the equivalence of the two gauges by giving the explicit gauge transformation for the massive scalar wave solutions. This result is also generalized to theories in which a scalar field
can have modified dispersion relations.

The paper is organized as follows.
In Sec. \ref{s:1} we recount the massive Branse--Dicke theory.
In Sec. \ref{s:2} we investigate the detector response in the conformal gauge and
in Sec. \ref{s:3} in the synchronous gauge.
In Sec. \ref{s:4} we give the angular and frequency characteristics of the one--arm one--way
detector.
In the Appendix \ref{A2} we derive the vacuum plane wave solution in the
linearized massive Brans--Dicke theory working directly in the synchronous gauge;
in the Appendix \ref{A3} we show the equivalence of the two gauges by explicitly giving the gauge transformation.

Greek indices $\mu,\nu,\ldots$ run from $0$ to $3$, Latin indices denote spatial coordinates:
$i,j,k,\ldots = 1,2,3$; $(x^0,x^1,x^2,x^3)$ are denoted also as $(t,x,y,z)$; colon "$:$" denotes contraction of tensors.

\section{Massive Brans-Dicke gravity}
\label{s:1}

In this section we give a brief account of the theory in which the gravitational interaction is mediated by two fields,
the standard metric tensor field and the massive scalar field. We rederive solutions of the linearized field equations
for the general gravitational wave comprising two massless tensor modes and one massive scalar mode.

We recall that a class of massive scalar--tensor theories described by the action
\begin{eqnarray}
\label{eq:BDaction}
S[g_{\mu\nu},\phi,\psi_m] & = & S_{g}[g_{\mu\nu},\phi] + S_{m}[g_{\mu\nu},\psi_{\rm m}], \\
S_{g}[g_{\mu\nu},\phi] & = & \frac{1}{16\pi}\int d^4x\,\sqrt{-g}\left[ \phi\,R -
\frac{\omega(\phi)}{\phi}\partial^{\mu}\phi\partial_{\mu}\phi + V(\phi) \right], \\
S_{m}[g_{\mu\nu},\psi_{\rm m}] & = & \int d^4x\,\sqrt{-g}L_{\rm m}[\psi_{\rm m},g_{\mu\nu}],
\end{eqnarray}
where $g_{\mu\nu}$ is a metric field, $\phi$  is a scalar field and $\psi_{\rm m}$ denotes collection of matter fields were introduced in \cite{Berg68}, \cite{Wag70}. Here $\omega$ and $V$ are two coupling functions and $R$ is the scalar curvature of the metric. The effects of the function $\omega$ on the dynamics of compact binaries have been extensively studied in \cite{WillZag89}, \cite{DEF92}, \cite{DEF96a}, \cite{DEF96b} or in the cosmological context in \cite{NaChiSu02}. The self--interaction potential $V$ in turn can play a role of the cosmological constant and give rise a mass term in the linearized theory.

In what follows we consider the massive Brans-Dicke theory in which the coupling parameter $\omega$ is given by the constant Brans--Dicke parameter, $\omega(\phi)=\omega_{BD}$. The field equations obtained by varying the action $S[g_{\mu\nu},\phi]$
with respect to the metric $g_{\mu\nu}$ and the scalar field $\phi$ are given by \cite{FM2003}
\begin{eqnarray}
\label{eq:feq:BDM}
R_{\mu\nu} - \frac12g_{\mu\nu}R - \frac{V(\phi)}{\phi }& = & \frac{8\pi T_{\mu\nu}}{\phi} +
\frac{\omega_{BD}}{\phi^2}\left( \phi_{,\mu}\phi_{,\mu}-\frac12 g_{\mu\nu}\phi_{,\alpha}{}^{,\alpha} \right)
+ \frac{\phi_{,\mu\nu} - \Box_{g}\phi}{\phi}, \\
\Box_{g}\phi + \frac{\phi V'(\phi) - 2V(\phi)}{3+2\omega_{BD}} & = & \frac{8\pi T}{3+2\omega_{BD}}  \nonumber,
\end{eqnarray}
where $R_{\mu\nu}$ is the Ricci tensor, $\Box_{g}\equiv (-g)^{-1/2}\partial_{\mu}(-g)^{1/2}g^{\mu\nu}\partial_{\nu}$, $T^{\mu\nu}\equiv\frac{2}{\sqrt{-g}}\frac{\delta S}{\delta g_{\mu\nu}}$, $T\equiv g_{\mu\nu}T^{\mu\nu}$ and
$V'\equiv\frac{dV}{d\phi}$. We consider small perturbations over the background configuration of the Minkowski metric
$\eta_{\mu\nu}$ and a constant field $\phi(x)=\phi_{0}$,
\begin{eqnarray}
\label{eq:gp:lin}
& g_{\mu\nu} = \eta_{\mu\nu} + h_{\mu\nu}, \quad \phi = \phi_{0} + \delta\phi,
\qquad |h_{\mu\nu}|\ll 1,\qquad|\delta\phi|\ll 1 &.
\end{eqnarray}
To preserve the asymptotic flatness of solutions and to neglect higher--order self--interaction terms for the scalar field
beside the mass term we assume \cite{ABWZ12} $V(\phi)=\frac12V''(\phi_0)\delta\phi^2$.
Substituting (\ref{eq:gp:lin}) into the field equations (\ref{eq:feq:BDM}) and introducing the mass of the scalar field
$m^2\equiv-\frac{\phi_0}{3+2\omega_{BD}}V''(\phi_0)$ one finds
\begin{eqnarray}
\label{eq:ein:lin}
& R_{\mu\nu}^{(1)} - \frac12g_{\mu\nu}R^{(1)}
 =  -\Phi_{,\mu\nu} + \eta_{\mu\nu}\Box_{\eta}\Phi & \\
\label{eq:BD:lin}
& \Box_{\eta}\Phi  =  m^2\Phi, &
\end{eqnarray}
where $\Box_{\eta}\equiv\eta^{\mu\nu}\partial_{\mu}\partial_{\nu}$, $\Phi\equiv-\frac{\delta\phi}{\phi_{0}}$
and we denote $R_{\mu\alpha\nu\beta}^{(1)}$, $R_{\mu\nu}^{(1)}$ and $R^{(1)}$
the linearizations of the Riemann tensor, Ricci tensor and Ricci scalar to first order in $h_{\mu\nu}$ respectively
(the explicit form of the $R^{(1)}_{\mu\alpha\nu\beta}$ is given in \cite{MTW} and recalled in the Appendix \ref{A2}).

One way to obtain the solutions of the linearized field equations (\ref{eq:ein:lin}), (\ref{eq:BD:lin})
(see e.g.\cite{MagNic00}) is to define
\begin{eqnarray}
\label{eq:theta}
\theta_{\mu\nu} & \equiv & h_{\mu\nu} - \eta_{\mu\nu}\left(\frac12 h - \Phi\right),\qquad h=\eta^{\mu\nu}h_{\mu\nu}
\end{eqnarray}
and to use the gauge freedom $h_{\mu\nu}\rightarrow h'_{\mu\nu}$, $\phi\rightarrow\phi'$,
\begin{eqnarray}
\label{eq:gfreedom}
h'_{\mu\nu}(x) & = &  h_{\mu\nu}(x) - \zeta_{(\mu,\nu)}, \qquad |\zeta_{\mu}|\ll 1 \\
\phi'(x) & = &  \phi(x), \nonumber
\end{eqnarray}
with the gauge parameter $\zeta_{\mu}$ satisfying
\begin{eqnarray}
\Box_{\eta}\zeta_{\mu}=\theta_{\mu\nu}{}^{,\nu},
\end{eqnarray}
to impose on $\theta_{\mu\nu}$ the Lorentz gauge condition (we omit primes in the transformed fields)
\begin{eqnarray}
\label{eq:glor}
\theta^{\mu\nu}{}_{,\nu} & = & 0.
\end{eqnarray}
In this gauge the field equations have the form of the wave and the Klein--Gordon equations in the flat spacetime,
\begin{eqnarray}
\label{eq:theta:lor}
\Box_{\eta}\theta_{\mu\nu} & = & 0,  \\
\label{eq:BD:lin:lor}
\Box_{\eta}\Phi & = & m^2\Phi,
\end{eqnarray}
describing the (superpositions) of the plane monochromatic waves
\begin{eqnarray}
\label{eq:theta:sol1}
\theta_{\mu\nu} & = & A_{\mu\nu}\,e^{-ik_{\mu}x^{\mu}},\quad k^\mu = (\omega,{\bf k}), \;\;\; \omega=|{\bf k}|, \quad k_{\mu}A^{\mu\nu}=0  \\
\label{eq:BD:lin:sol1}
\Phi & = & A\,e^{-il_{\mu}x^{\mu}}, \quad \;\;\;\;\; l^\mu = (\omega,{\bf l}), \;\;\;\; \omega=\sqrt{{\bf l}^2+m^2}
\end{eqnarray}
which can be written as
\begin{eqnarray}
\label{eq:theta:sol2}
\theta_{\mu\nu} & = & A_{\mu\nu}\,e^{i\omega (t - \bO\cdot{\bf x})}, \qquad \bO={\bf k}/|{\bf k}| \\
\label{eq:BD:lin:sol2}
\Phi & = & A\,e^{i\omega (t - \frac{\bO\cdot{\bf x}}{{\rm v}(\omega)})}, \qquad \;\;\; \bO={\bf l}/|{\bf l}|,
\end{eqnarray}
where $\bO$'s are unit vectors along the wave propagation and ${\rm v}$ is the $\omega$--dependent phase velocity,
${\rm v}(\omega)=\frac{|\omega|}{\sqrt{\omega^2-m^2}}$, of the scalar field.
(The phase velocity diverges when $\omega$ tends to $m$. This is because the wavelength $\lambda=2\pi/\sqrt{\omega^2-m^2}$ grows then to infinity; in the limit $\omega=m$ the solutions of
Eqs. (\ref{eq:theta:sol2}), (\ref{eq:BD:lin:sol2}) are therefore proportional to space--independent oscillations $e^{imt}$.)
At this step we have $h_{\mu\nu}=\theta_{\mu\nu}-\eta_{\mu\nu}(\frac12\,\theta-\Phi)$, where $\theta=\eta^{\mu\nu}\theta_{\mu\nu}$ but the Lorentz condition (\ref{eq:glor}) is preserved under the supplementary gauge transformation (\ref{eq:gfreedom}) with $\zeta_{\mu}$ satisfying
\begin{eqnarray}
\Box_{\eta}\zeta_{\mu} & = & 0, \\
\zeta^{\mu}{}_{,\mu} & = & -\frac12 \theta
\end{eqnarray}
rendering the trace of $\theta_{\mu\nu}$ equal to zero and giving $h_{\mu\nu}=\theta_{\mu\nu}+\Phi\,\eta_{\mu\nu}$.
The residual gauge freedom
\begin{eqnarray}
\Box_{\eta}\zeta_{\mu} & = & 0, \\
\label{eq:tr0}
\zeta^{\mu}{}_{,\mu} & = & 0
\end{eqnarray}
is exactly the same as the gauge freedom that is left after specifying the Lorentz condition and imposing traceless condition on the metric perturbation in GR and can be used to transform $\theta_{\mu\nu}$ to the transverse--traceless (TT) form \cite{MTW}. We call the obtained gauge the {\it conformal gauge} since it allows to represent a gravitational wave as the sum
\begin{eqnarray}
\label{eq:zeta:res}
h_{\mu\nu}(t,{\bf x}) & = & A_{\mu\nu}(t,{\bf x}) + \Phi(t,{\bf x})\eta_{\mu\nu},
\end{eqnarray}
of the TT wave $A_{\mu\nu}$ satisfying $A_{\mu0}=0$, $A^{ij}{}_{,j}=0$, $A^{i}{}_{i}=0$
and the scalar wave conformal to the Minkowski metric, $\Phi(t,{\bf x})\eta_{\mu\nu}$. For the plane wave propagating
in the $-z$ direction (\ref{eq:zeta:res}) simplifies to
\begin{eqnarray}
\label{eq:c:z}
h_{\mu\nu}(t,z) & = & A^{+}(t+z)\epsilon^{+}_{\mu\nu} + A^{\times}(t+z)\epsilon^{\times}_{\mu\nu} + \Phi(t,z)\epsilon^{s}_{\mu\nu},
\end{eqnarray}
with the polarization tensors
\begin{eqnarray}
\label{eq:polar:pc}
\epsilon^{s} & = &
\left(
\begin{array}{cccc}
-1 & 0 & 0 & 0 \\
0 & 1 & 0 & 0 \\
0 & 0 & 1 & 0 \\
0 & 0 & 0 & 1
\end{array}
\right),
\qquad
\epsilon^{+} = \left(
\begin{array}{cccc}
0 & 0 & 0 & 0 \\
0 & 1 & 0 & 0 \\
0 & 0 & -1 & 0 \\
0 & 0 & 0 & 0
\end{array}
\right),
\qquad
\epsilon^{\times} = \left(
\begin{array}{cccc}
0 & 0 & 0 & 0 \\
0 & 0 & 1 & 0 \\
0 & 1 & 0 & 0 \\
0 & 0 & 0 & 0
\end{array}
\right).
\end{eqnarray}
In the Appendix \ref{A2} we derive the solutions to the linearized vacuum field equations (\ref{eq:ein:lin}), (\ref{eq:BD:lin}) directly in the gauge $h_{\mu0}=0$.

\section{Detector response in the conformal gauge}
\label{s:2}

In this section we investigate the motion of a free test mass in the background of the plane, massive scalar
gravitational wave in the conformal gauge. The result will enable us to obtain the response of the laser interferometer. In what follows we make use of the derivation given in \cite{Shibata94}, \cite{Nakao01} for the massless scalar field.

To this end we first consider an arbitrary conformal wave moving in the direction $-z$ with the velocity
${\rm v}$, ${\rm h}_{s}(t,z)=h_{s}(t+z/{\rm v})$ (so e.g. it can be one of a Fourier modes of the
Eq. (\ref{eq:BD:lin:sol2})), in which case the background geometry has the form:
\begin{eqnarray}
\label{eq:conf}
ds^2 & = & \left[1 + h_{s}\left(t + \frac{z}{{\rm v}}\right)\right]\eta_{\mu\nu}dx^{\mu}dx^{\nu}.
\end{eqnarray}
Free motion of a test body can be obtained from the Lagrangian
$$
L({\bf x},\dot{\bf x})= \frac12\left[1 + h_{s}\left(t + \frac{z}{{\rm v}}\right)\right]\left( -\dot{t}^2 + \dot{x}^2 + \dot{y}^2 + \dot{z}^2 \right),
$$
where the dot stands for the proper time derivative, $\dot{x}^{\mu}(\tau)\equiv \frac{d x^{\mu}(\tau)}{d \tau}$.
The equations of motion read
\begin{eqnarray}
\label{eq:eomt}
\frac{d}{d\tau}\left[ (1 + h_{s})\dot{t} \right] & = & -\frac12 {\rm h}_{s,t}\left( -\dot{t}^2 + \dot{x}^2 + \dot{y}^2 + \dot{z}^2 \right)\\
\label{eq:eomi}
\frac{d}{d\tau}\left[ (1 + h_{s})\dot{x}^{i} \right] & = & \frac12 {\rm h}_{s,i}\left( -\dot{t}^2 + \dot{x}^2 + \dot{y}^2 + \dot{z}^2 \right),\qquad i=1,2,3.
\end{eqnarray}
We assume that in the absence of gravitational wave the test body was at rest with respect to the coordinates and hence we seek for the leading order solution having the form $\dot{t}=1 + A_{t}h$, $\dot{x}^i= A_{i}h$. From Eqs. (\ref{eq:eomi}) it follows that, as a first step, we can specify $A_x=A_y=0$. Using the (leading order) identities ${\rm h}_{s,t}=\dot{h}_{s}$ and ${\rm h}_{s,z}=\dot{h}_{s}/{\rm v}$ from Eqs. (\ref{eq:eomt}), (\ref{eq:eomi}) it then follows that
$A_{t} = -\frac12$, $A_{z}=-\frac{1}{2 {\rm v}}$ and we get
\begin{eqnarray}
\label{eq:sol1x}
x(t) & = & x_{0} \\
\label{eq:sol1y}
y(t) & = & y_{0} \\
\label{eq:sol1z}
z(t) & = & z_{0} - \frac{1}{2 {\rm v}}\int\limits_{\infty}^{t}h_{s}[t'+ z(t')/{\rm v}]\,dt' = z_{0} - \frac{1}{2 {\rm v}}\int\limits_{\infty}^{t+\frac{z_{0}}{{\rm v}}}h_{s}(v)\,dv \\
\label{eq:sol1t}
\tau(t) & = & t + \frac12\int\limits_{\infty}^{t+\frac{z_{0}}{{\rm v}}}h_{s}(v)\,dv,
\end{eqnarray}
where $v(t')=t'+z(t')/{\rm v}$ and $x_0$, $y_0$, $z_0$ are the initial positions set at $t\rightarrow-\infty$.
In order to have the motion of the test particle in the case of the plane wave propagating along the unit vector
$\bO=(-\cos{\phi}\sin{\theta},-\sin{\phi}\sin{\theta},-\cos{\theta})$ one rotates the frame
\begin{eqnarray}
\label{eq:rot1}
\left(
\begin{array}{c}
x_{new} \\
y_{new} \\
z_{new}
\end{array}
\right) & = &
\left(
\begin{array}{ccc}
\cos{\theta}\cos{\phi} & -\sin{\phi} & \sin{\theta}\cos{\phi} \\
\cos{\theta}\sin{\phi} & \cos{\phi} & \sin{\theta}\sin{\phi} \\
-\sin{\theta} & 0 & \cos{\theta}
\end{array}
\right)
\left(
\begin{array}{c}
x_{} \\
y_{} \\
z_{}
\end{array}
\right).
\end{eqnarray}
to get in the new coordinates (we omit the subindex '$new$'):
\begin{eqnarray}
\label{eq:solvec}
{\bf x}(t) & = & {\bf x}_{0} + \frac{\bO}{2 {\rm v}}\int\limits_{-\infty}^{t-\frac{\bO\cdot{\bf x}_{0}}{{\rm v}}}h_{s}(v)\,dv \\
\label{eq:solt}
\tau(t) & = & t + \frac12\int\limits_{-\infty}^{t-\frac{\bO\cdot{\bf x}_{0}}{{\rm v}}}h_{s}(v)\,dv,
\end{eqnarray}
with ${\bf x}_{0}=(x_{0},y_{0},z_{0})$ and
$\bO\cdot{\bf x}_{0} = -x_{0}\cos{\phi}\sin{\theta} - y_{0}\sin{\phi}\sin{\theta} - z_{0}\cos{\theta}$.

To obtain the detector response we place the freely moving  emitter and the detector initially at the points ${\bf x}_{E0}$ and ${\bf x}_{D0}={\bf x}_{E0} + L\,{\bf n}_{ED}$ respectively, where ${\bf n}_{ED}$ is the unit vector
from the emitter to the detector and $L$ is the length of the detector arm both defined with respect to the
$3$--dimensional Euclidean metric $\delta_{ij}$. We assume that clocks that measure the proper times along the trajectories of the emitter and detector were synchronized in the absence of the wave. Then the {\it time--of--flight} of the light signal from $E$ to $D$, $\Delta \tau_{ED}(t) \equiv \tau_{D}(t) - \tau_{E}\left[t-\delta t(t)\right]$, where $t$ and $t-\delta t(t)$ are the coordinate times of the emission and the detection respectively, is a coordinate--independent quantity, in fact it enters the detector response of all the gravitational wave laser interferometers. Namely, if we compare a laser signal $A_{E}=A_{L}e^{i\omega_{L}\tau_{E}}$ with angular frequency $\omega_{L}$ sent from the emitter to the detector with the identical template laser signal $A_{D}=A_{L}e^{i\omega_{L}\tau_{D}}$ at the detector
the change in the phase will be proportional to $\Delta\tau_{ED}$:
\begin{eqnarray}
\label{eq:laser_1}
A_{D}(\tau_{D}(t)) - A_{E}\left[\tau_{E}(t-\delta t(t))\right] \simeq A_{L}\,i\omega_{L}\left[ \tau_{D}(t) - \tau_{E}[t-\delta t(t)] \right].
\end{eqnarray}
In the background given by Eq. (\ref{eq:conf}) light travels along the null lines of the Minkowski metric
$\eta_{\mu\nu}$ thus to the leading order
\begin{eqnarray}
\delta t(t) & = & |{\bf x}_{D}(t) - {\bf x}_{E}(t-L)| =
|\,{\bf x}_{D0} - {\bf x}_{E0} + \frac{\bO}{2 {\rm v}}
\int\limits_{t-L-\frac{\bO\cdot{\bf x}_{E0}}{{\rm v}}}^{t-\frac{\bO\cdot{\bf x}_{D0}}{{\rm v}}}h_{s}(v)\,dv\,| \nonumber \\
& \simeq &
L + \frac{\bO\cdot{\bf n}_{ED}}{2 {\rm v}}
\int\limits_{t-L-\frac{\bO\cdot{\bf x}_{E0}}{{\rm v}}}^{t-\frac{\bO\cdot{\bf x}_{D0}}{{\rm v}}}h_{s}(v)\,dv \nonumber
\end{eqnarray}
and using (\ref{eq:solt})
\begin{eqnarray}
\label{eq:DT1}
\Delta \tau_{ED}(t)
\label{eq:tau}
& \simeq &
t + \frac12\int\limits_{\infty}^{t-\frac{\bO\cdot{\bf x}_{D0}}{{\rm v}}}h_{s}(v)\,dv \\
&&
- \left[ t - L - \frac{\bO\cdot{\bf n}_{ED}}{2 {\rm v}}
\int\limits_{t-L-\frac{\bO\cdot{\bf x}_{E0}}{{\rm v}}}^{t-\frac{\bO\cdot{\bf x}_{D0}}{{\rm v}}}h_{s}(v)\,dv
+ \frac12\int\limits_{\infty}^{t-L-\frac{\bO\cdot{\bf x}_{E0}}{{\rm v}}}h_{s}(v)\,dv\right] \nonumber \\
& = &
L + \frac12\left(1 + \frac{\bO\cdot{\bf n}_{ED}}{{\rm v}}\right)
\int\limits_{t-L-\frac{\bO\cdot{\bf x}_{E0}}{{\rm v}}}^{t - \frac{\bO\cdot{\bf x}_{D0}}{{\rm v}}}h_{s}(v)\,dv. \nonumber
\end{eqnarray}
In Eq.(\ref{eq:sol1z}) we have parametrized the constant phase surfaces $v$'s with the coordinate time $t'$ along the particle trajectory. Now we change the parametrization and will use a parameter $\lambda$ along the laser ray.
To this end we notice that the trajectory $[{\rm t}_{0}(\lambda),{\bf x}_{0}(\lambda)]$ of the laser ray from ${\bf x}_{E0}$ to ${\bf x}_{D0}$ (end points are needed only in the $0$-th order in the argument of the integrand $h_{s}$ of Eq. (\ref{eq:tau})) is given by
\begin{eqnarray}
\label{eq:rays}
{\rm t}_{0}(\lambda) = \lambda, \qquad {\rm \bf x}_{0}(\lambda) = {\rm \bf x}_{D0} - {\bf n}_{ED}(t-\lambda).
\end{eqnarray}
Thus, changing the variables, $v(\lambda) = {\rm t}_{0}(\lambda) - \bO\cdot{\rm \bf x}_{0}(\lambda)/{\rm v}$, one finds
\begin{eqnarray}
\label{eq:DT3a}
\Delta \tau_{ED}(t)
& = &
L + \frac{1 - \left(\frac{\bO\cdot{\bf n}_{ED}}{{\rm v}}\right)^2}{2} \int\limits_{t-L}^{t}h_{s}\left[v(\lambda)\right]\,d\lambda \\
\label{eq:DT3b}
& = &
L + \frac{1 - \left(\bO\cdot{\bf n}_{ED}\right)^2  + (1 - \frac{1}{{\rm v}^2})\left(\bO\cdot{\bf n}_{ED}\right)^2}{2} \int\limits_{t-L}^{t}h_{s}\left[v(\lambda)\right]\,d\lambda \nonumber \\
\label{eq:DT3c}
& = &
L + \frac{{\bf n}_{ED}\otimes{\bf n}_{ED}:\left[\be^{st}  + (1 - \frac{1}{{\rm v}^2})\be^{sl}\right] }{2} \int\limits_{t-L}^{t}h_{s}\left[v(\lambda)\right]\,d\lambda, \nonumber
\end{eqnarray}
where the $3\times3$ matrices, $\be^{st}$ and $\be^{sl}$, are defined in the Appendix \ref{A1}.
They play the role of the two (spatial) polarization tensors of the scalar transversal and
scalar longitudinal modes whose components in the synchronous gauge and in the source frame read
\begin{eqnarray}
\label{eq:polarizations}
\epsilon^{st} & = &
\left(
\begin{array}{ccc}
1 & 0 & 0 \\
0 & 1 & 0 \\
0 & 0 & 0
\end{array}
\right),
\qquad
\epsilon^{sl} = \left(
\begin{array}{ccc}
0 & 0 & 0 \\
0 & 0 & 0 \\
0 & 0 & 1
\end{array}
\right).
\end{eqnarray}

\section{Detector response in the synchronous gauge}
\label{s:3}

In the synchronous gauge the metric is given by
\begin{eqnarray}
\label{eq:b:synch}
{\rm \bf g} = -dt^2 + \left(\delta_{ij} + {\rm h}_{ij}\right)dx^i\,dx^j.
\end{eqnarray}
The condition ${\rm h}_{\mu0}=0$ implies that the motion of test bodies is trivial:
one can assume that the emitter and detector stay at fixed coordinates, $x^{i}(t)=const.$ and the coordinate time $t$ is equal to the proper times along their trajectories, $\tau_{E}(t)=\tau_{D}(t)=t$. The time of flight is then given by the difference in the coordinate times of the detection and emission and is determined by the integral
\begin{eqnarray}
\label{eq:tof_T1}
\Delta \tau_{ED}(t) = {\rm t}(\lambda_{D}) - {\rm t}(\lambda_{E}) =
\int\limits_{\gamma_{ED}}\sqrt{{\rm g}_{ij}\left[\rm{t}(\lambda),{\rm \bf x(\lambda)}\right]\frac{d{\rm x}^{i}}{d\lambda}\frac{d{\rm x}^{j}}{d\lambda}}\,d\lambda,
\end{eqnarray}
where $[{\rm t}(\lambda),\gamma_{ED}(\lambda)]=[{\rm t}(\lambda),{\rm \bf x}(\lambda)]$ parametrizes the null geodesic of the light ray from the emitter to the detector with the arrival time at the detector ${\rm t}(\lambda_{D})=t$. In the synchronous coordinates trajectories of photons in the perturbed background (\ref{eq:b:synch}) differ in general from straight lines but as was shown by Finn \cite{Finn09} (see also \cite{AB2012},\cite{TA10}) the integral on the right hand side of Eq.(\ref{eq:tof_T1}) is unchanged when
computed along the {\it unperturbed} trajectory
\begin{eqnarray}
\label{eq:unpert_tr}
{\rm t}_{0}(\lambda) = \lambda, \qquad {\rm \bf x}_{0}(\lambda) = {\bf x}_{D} - {\bf n}_{ED}(t - \lambda).
\end{eqnarray}
This gives to the leading order
\begin{eqnarray}
\label{eq:tof_T2a}
\Delta \tau_{ED}(t) & = &
L + \frac{{\bf n}_{ED}\otimes{\bf n}_{ED}}{2}:
\int\limits_{t-L}^{t}{\rm \bf h}\left[\rm{t}_{0}(\lambda),{\rm \bf x}_{0}(\lambda)\right]\,d\lambda \\
\label{eq:tof_T2b}
& = &
L + \frac{{\bf n}_{ED}\otimes{\bf n}_{ED}:\be^{p}}{2}\int\limits_{t-L}^{t}h_{p}\left[v(\lambda)\right]\,d\lambda, \nonumber
\end{eqnarray}
where in the last line of Eq.(\ref{eq:tof_T2a}) we have specified the $p$--polarized plane wave having the (spatial) polarization tensor $\be^{p}$ propagating along the vector $\bO$ with the velocity ${\rm v}$,
${\rm \bf h}(t,{\bf x})=\be^{p}\,h_{p}(t-\bO\cdot{\bf x}/{\rm v})$, and we have defined
$v(\lambda) = {\rm t}_{0}(\lambda) - \bO\cdot{\rm \bf x}_{0}(\lambda)/{\rm v}$.

Comparison of Eqs.(\ref{eq:tof_T2a}) and (\ref{eq:DT3a}) suggests that the scalar wave defined in the conformal gauge corresponds to a superposition of scalar transversal and scalar longitudinal waves defined in the synchronous gauge (and in the frame in which the wave propagates in the $-z$ direction) as:
\begin{eqnarray}
\label{eq:synch}
ds^2 = -dt^2 + h_{st}\left(t + \frac{z}{{\rm v}}\right)(dx^2 + dy^2) + h_{sl}\left(t + \frac{z}{{\rm v}}\right)dz^2,
\end{eqnarray}
with
\begin{eqnarray}
\label{eq:hsl:hst:rel}
h_{st} = h_{s},\qquad h_{sl}=\left(1-\frac{1}{{\rm v}^2}\right)h_{s}
\end{eqnarray}
This is indeed the case and in the Appendix \ref{A3} we give the gauge transformation that connects both gauges. There it is shown that more general result holds: given a scalar mode $h_{s}(t,z)$ that in the conformal gauge satisfies the
field equation $\partial_{t}^2h_{s} = F[\partial_{z}]h_{s}$ the corresponding synchronous gauge wave is given by
\begin{eqnarray}
\label{eq:gen}
{\rm \bf h}(t,z) & = &
\text{diag}\left(0, h_s(t,z), h_s(t,z), \frac{F\left[\partial_{z}\right] - \partial_{z}^2}{F\left[\partial_{z}\right]}h_{s}(t,z)\right).
\end{eqnarray}
Beside the massive Brans--Dicke and $f(R)$ theories this result includes also theories with modified dispersion relation \cite{MiYuW11} that can arise e.g. as the effective level of some approaches to quantum gravity or in Lorentz--symmetry violating theories \cite{GAC}, \cite{HL09}, \cite{Jac07}. We note here that usually the effects of modification of the dispersion relations are considered to be suppressed by the Planck scale but \cite{CPSUV04}, \cite{CPS06} indicated the possibility of their enhancement when the renormalization is taken into an account in Lorentz symmetry breaking models.

\section{Sensitivity of one-arm interferometer}
\label{s:4}

Integration of the right hand side Eq. (\ref{eq:tof_T2a}) for the scalar monochromatic $p$--polarized wave
${\rm h}(t.{\bf x})=h_{p}^{0}\,e^{i\,\omega(t - \frac{\bO\cdot{\bf x}}{{\rm v}})}$ ($p=sl,st$) leads to the following
result for the time--dependent part of the time of flight:
\begin{eqnarray}
\label{eq:sED1}
s_{ED}(t) & \equiv & \frac{\Delta\tau_{ED}(t)}{T} = F^{p}({\bf n}_{ED})\,{\cal T}(x;c_{ED})\,{\rm h}(t,{\bf x}_{D}),
\end{eqnarray}
where $c_{ED}\equiv\frac{\bO\cdot{\bf n}_{ED}}{{\rm v}}$, $x\equiv L\,\omega$,
$F^{p}({\bf n}_{ED}):=\frac{1}{2}{\bf n}_{ED}\otimes{\bf n}_{ED}:\be^{p}$
is the (one--arm) antenna pattern function and
${\cal T}(x;c_{ED})\equiv\sinc\left[ \frac{x(1-c_{ED})}{2} \right]\,e^{-i\frac{x(1-c_{ED})}{2}}$
is the frequency response of the (one--arm and one--way) detector.

\begin{figure}[htp]
\begin{center}
\subfigure{
$a$
\label{f:TT2a}
\includegraphics[width=9pc]{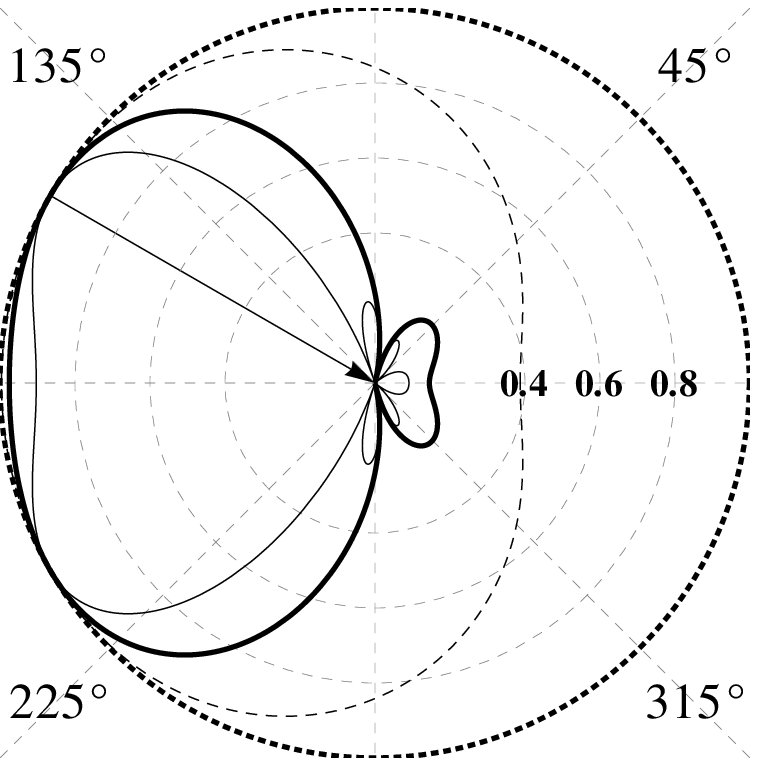}}
\hspace{2cm}
\subfigure{
$b$
\label{f:TT2b}
\includegraphics[width=9pc]{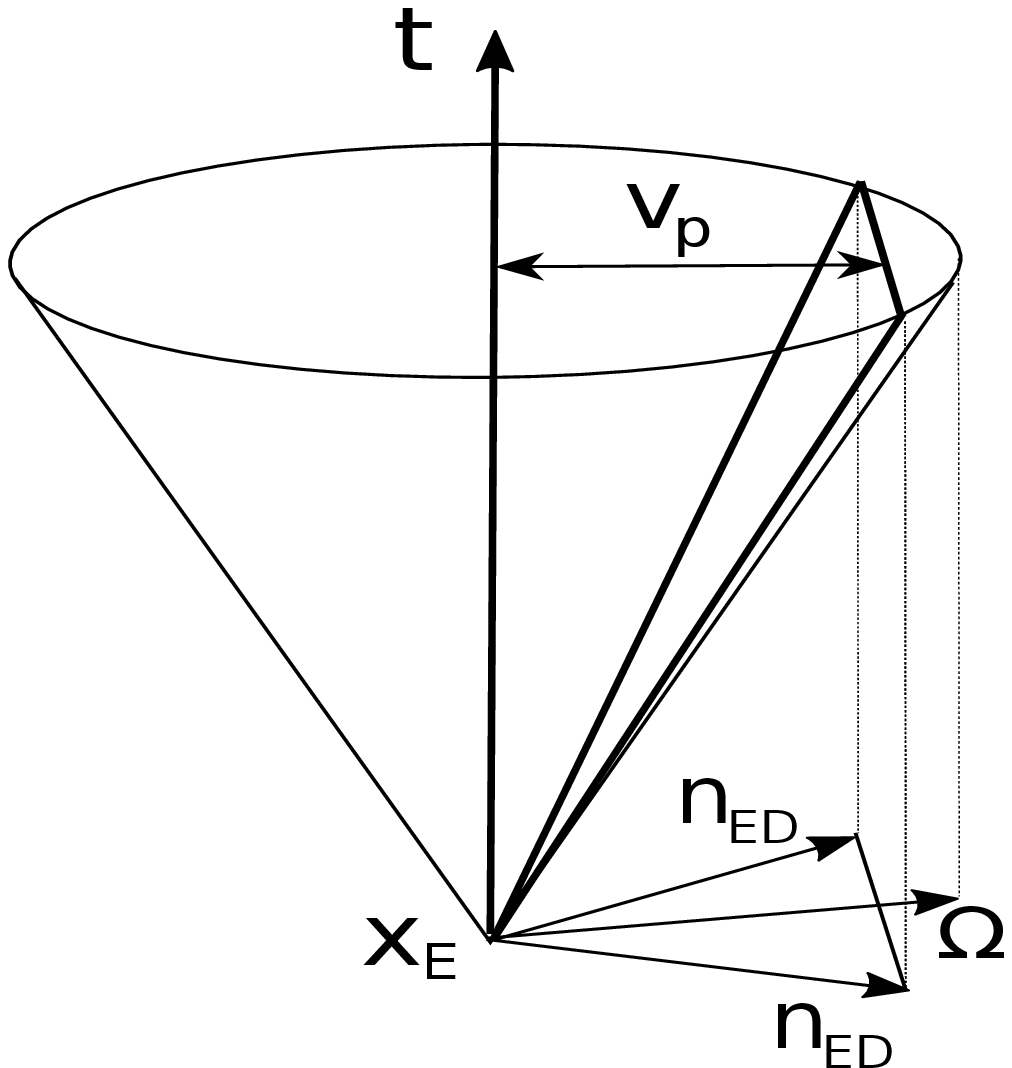}}
\end{center}
\caption{Left: plane section of the frequency transfer function $|{\cal T}|$. The arm ${\bf n}_{ED}$ of the detector is inclined at
$0^o$, ${\rm v}=\bO\cdot{\bf n}_{ED}=\cos{\frac{\pi}{6}}$; the arrow represents one of the directions $\bO_{\max}$ of the maximal
sensitivity. The normalized frequencies $x=\omega\,L$ are equal $0$ (dotted), $2$ (dashed), $5$ (thick) and $10$ (thin).
Right: section of the null cone and the hypersurface of the constant phase passing through the emitter.}
\end{figure}

\begin{figure}[htp]
\begin{center}
\subfigure{
$a$
\label{f:st}
\includegraphics[width=10pc]{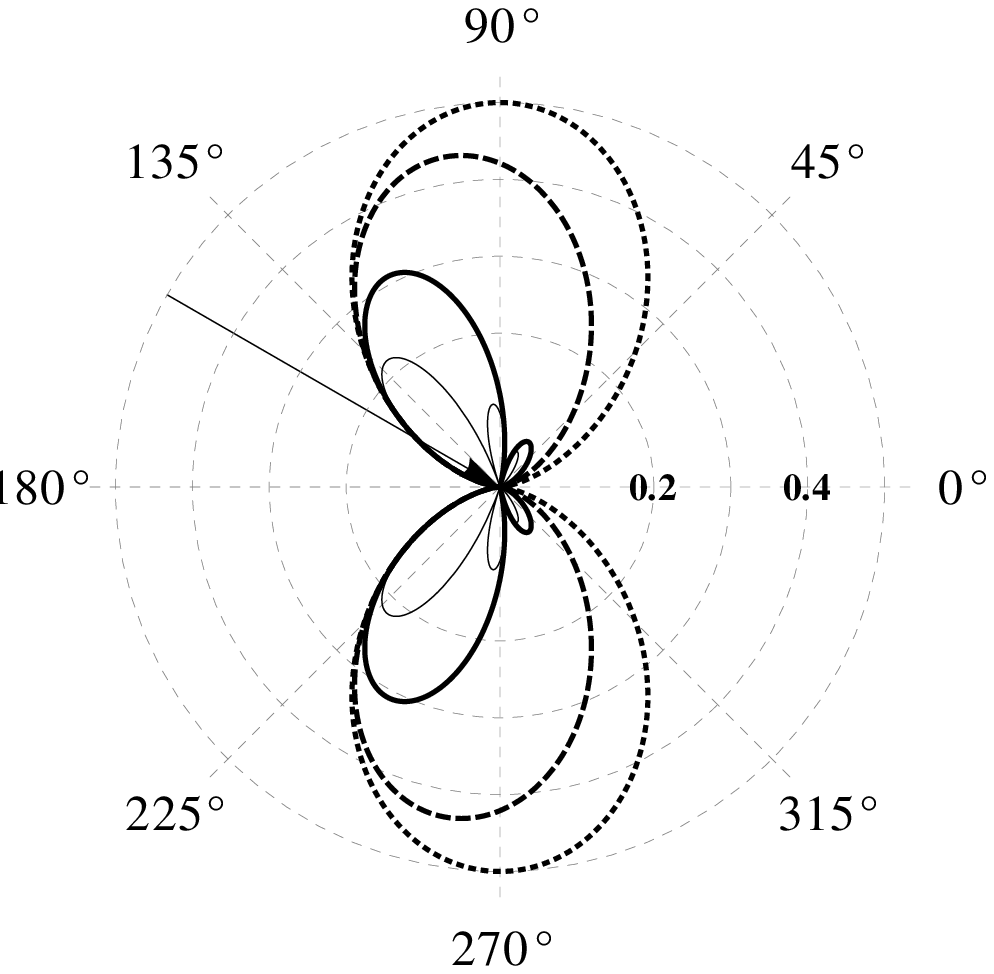}}
\hspace{0.1cm}
\subfigure{
$b$
\label{f:sl}
\includegraphics[width=11pc]{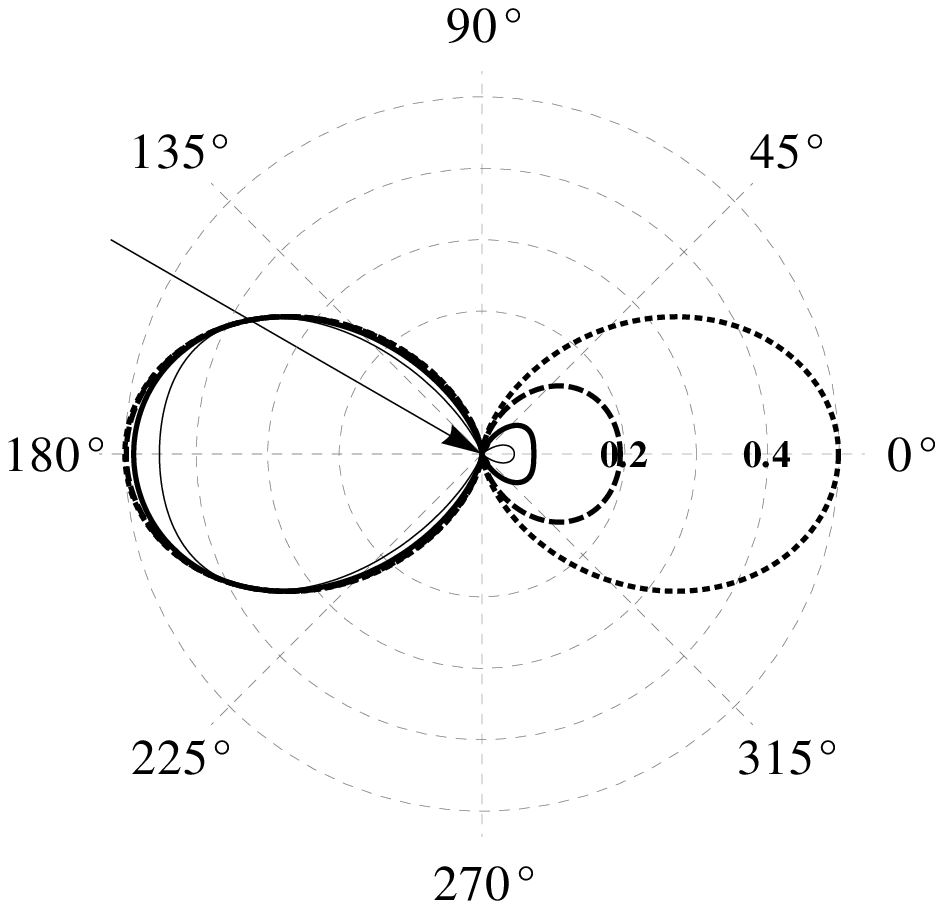}}
\hspace{0.1cm}
\subfigure{
$c$
\label{f:s}
\includegraphics[width=11pc]{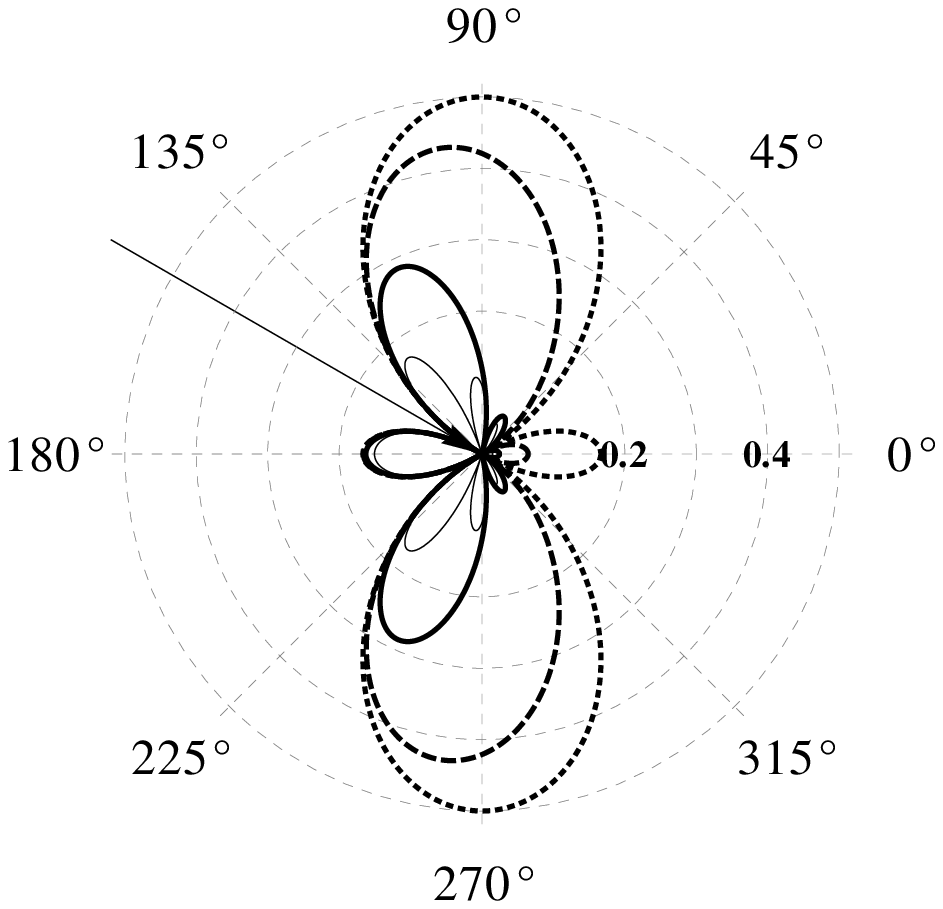}}
\end{center}
\caption{Planar sections of the detector response $|\,F^{p}({\bf n}_{ED})\,{\cal T}(x;c_{ED})\,|$ for the $st$ $(a)$,
$sl$ $(b)$ and $s$ $(c)$ modes; notation and parameters as in Fig.\ref{f:TT2a}. In cases $a$, $b$, $c$ responses for the
signal coming from $150^o$ are frequency--independent, for $c$ it is zero.}
\end{figure}

It was already observed in \cite{BPPP08} that the
frequency--independent maximum of the frequency transfer function, ${\cal T}=1$, is achieved when the wave passes through the arm at the particular angle $\bO\cdot{\bf n}_{ED}=\cos{\vartheta}={\rm v}$ and this is only possible when ${\rm v}\leq 1$; this is illustrated in Fig.\ref{f:TT2a}. To see the geometrical picture behind that angle we notice first that it is zero when ${\rm v}=1$. In this case as remarked in \cite{TA10},\cite{AB2012} the one-arm frequency transfer function ${\cal T}$ does not depend on frequency. This together with non-vanishing antenna pattern function along the detector's arm for the scalar longitudinal mode leads to the preferable detection feasibility at high frequencies for that mode. The reason for this is simple: the spacetime trajectory of the light ray lies on the $3$--dimensional hypersurface of the constant phase of the passing plane gravitational wave therefore photons perceive it as a constant field. The same is true for ${\rm v}<1$ but in this case planes of the constant phases of the gravitational wave are no longer tangent to null cones of the light rays. When ${\rm v}<1$ the light cone formed by light rays emitted from ${\bf x}_{E}$ and the plane of the constant phase passing through ${\bf x}_{E}$ intersect and the null lines of the intersection determine a set of null trajectories that satisfy $\bO\cdot{\bf n}_{ED}={\rm v}$, see Fig.\ref{f:TT2b}. For ${\rm v}>1$ light cones and planes of the constant phases do not intersect and the effect does not arise.
Interestingly for ${\rm v}<1$ the direction of the maximum has non-vanishing components parallel and orthogonal to ${\bf n}_{ED}$ thus one expects that all polarization modes of gravitational waves, transversal and longitudinal, will share the property of having frequency--independent one-arm response functions, $F^{p}({\bf n}_{ED})\,{\cal T}(x;c_{ED})$, in the directions determined by $\bO\cdot{\bf n}_{ED}={\rm v}$ (see Figs.\ref{f:st}, \ref{f:sl}).
In fact this feature for the $+$ and $\times$ tensorial modes was explored in \cite{BPPP08} and used in putting the limits on the speed of gravitational waves from pulsar timing and providing a bound on the graviton mass $m_{g}\leq 8.5\times 10^{-24}$eV. Here however we restrict ourselves from the beginning to the theories in which the standard $+$ and $\times$ modes travel with the speed of light unlike the scalar modes whose amplitudes as was shown in the previous chapter
are related. Thus the detector response for the scalar wave having in the conformal gauge the form
$\left[(1 + {\rm h}_{s}(t,{\bf x})\right]\eta_{\mu\nu}$ with the amplitude
${\rm h}_{s}(t,{\bf x})=h_{s}^{0}e^{i\,\omega(t-\bO\cdot{\bf x}/{\rm v})}$ is given by the Eq. (\ref{eq:sED1})
with the antenna pattern function that reads
\begin{eqnarray}
\label{eq:null:s}
F^{s}({\bf n}_{ED}) & = & F^{st}({\bf n}_{ED}) + \left(1 - \frac{1}{{\rm v}^2}\right)\,F^{sl}({\bf n}_{ED}) \\
& = &
\frac12-\frac{1}{{\rm v}^2}\,F^{sl}({\bf n}_{ED}) =
\frac{{\rm v}^2 - (\bO\cdot{\bf n}_{ED})^2}{2\,{\rm v}^2},\nonumber
\end{eqnarray}
if we use $F^{st} + F^{sl} = \frac12$, $F^{sl}=\frac12(\bO\cdot{\bf n}_{ED})^2$. We notice that
the antenna pattern function $F^{s}$ must preserve the property of frequency--independence for the detector response for the wave coming from the direction determined by $\bO\cdot{\bf n}_{ED}={\rm v}$ since both antenna patterns $F^{st}$ and $F^{sl}$ do. But interestingly for the particular combination of the scalar modes which is motivated on the theoretical ground by the massive scalar tensor theories the net sensitivity is null for $\bO\cdot{\bf n}_{ED}={\rm v}$,
(see Eq. (\ref{eq:null:s}) and Fig. \ref{f:s}).

Similarly one can obtain the detector response for the scalar wave that in the conformal gauge has the form
\begin{eqnarray}
{\rm h}_{s}(t,{\bf x}) & = &
h_{s}^{0}\,\be^{p}\cos{\left[\omega\left(t - \frac{\bO\cdot{\bf x}}{{\rm v}}\right) + \phi_{0}\right]};
\end{eqnarray}
it is given by (the real part of Eq.(\ref{eq:sED1}))
\begin{eqnarray}
\label{eq:sED2}
s_{ED}(t) & = &
\text{sinc}\left[ \frac{x(1-c_{ED})}{2} \right]\,F^{s}({\bf n}_{ED})\,
h_{s}^{0}\,\cos{\left[ \omega\left(t - \frac{\bO\cdot{\bf x}_{E}}{{\rm v}}\right) - \frac{x}{2}\left( 1+c_{ED} \right) + \phi_{0}\right]}.
\end{eqnarray}
We see that due to the relation (\ref{eq:hsl:hst:rel}) between the scalar transversal and longitudinal modes the initial phase $\phi_{0}$ of the $h^{st}$ and $h^{sl}$ signals is the same. From the one--arm response one can construct in a standard way other responses, e.g. the Michelson interferometer based on ${\bf n}_{ED_1}$ and ${\bf n}_{ED_2}$ is defined as
\begin{eqnarray}
\label{eq:M}
M(t) & = & s_{ED_1}(t-L) + s_{D_1E}(t) - s_{ED_2}(t-L) - s_{D_2E}(t).
\end{eqnarray}
Furthermore we see that the long wavelength (LW) limit defined as the leading $x$ term of the response (\ref{eq:sED2}) is given by
\begin{eqnarray}
\label{eq:sedlw}
s_{ED}^{LW}(t) & = & F^{s}({\bf n}_{ED})\,
h_{s}^{0}\,\cos{\left[ \omega\left(t - \frac{\bO\cdot{\bf x}_{E}}{{\rm v}}\right) + \phi_{0} \right]}
\end{eqnarray}
so the corresponding Michelson interferometer $M^{LW}$ will not discern $h^{s}$, $h^{st}$ and $h^{sl}$ signals whatever
orientation of the two arms ${\bf n}_{ED_1}$ and ${\bf n}_{ED_2}$ is (note that the same is true for some other responses e.g. the Sagnac interferometer $S(t)=s_{ED_1}(t-2L) + s_{D_1D_2}(t-L) + s_{D_2E}(t)- s_{ED_2}(t-2L) - s_{D_2D_1}(t-L) - s_{D_1E}(t)$).
However beyond the LW limit the responses of the  Michelson interferometers $M$ are different due to the orientation--dependent higher frequency terms which potentially enable to discriminate between the modes. On the other hand the difference between ${\rm v}=1$ and ${\rm v}\neq 1$ case shows up already in the LW limit, for instance, in the massive Brans-Dicke theory
as the corrections $\sim\bO\cdot{\bf x}_{E}\left(\frac{m}{\omega}\right)^2$
in the signal's phase.

\section{Summary}

The analysis of the response of the laser interferometer to passing gravitational waves is a starting point in the gravitational waves detection experiments. Especially at the present moment of awaiting the first detection by the advanced detectors there is a good opportunity to routinely confront GR with the alternative theories testing gravity in the new dynamical, relativistic regime \cite{IWMP15}.
The theoretical framework for classification of waves in alternative theories was given in \cite{Eardley73}
under the assumption of the minimal coupling of gravity to matter fields and with the restriction that the waves
must travel at exactly the speed of light. The present paper deals with theories where the former assumption is fulfilled but the later restriction is relaxed. The detectability of gravitational wave signals in the massless scalar--tensor theory of Brans and Dicke where both modes, scalar and tensor, move with the speed of light was studied in \cite{Will94} for inspiralling compact binaries. In \cite{Shibata94} the detectability of massless scalar waves was investigated in the case of gravitational collapse and the analysis of the detector response for those modes was also given. In this context finding the detection methods to discern between the scalar and tensor waves and thorough analysis of the signals in the detector was desirable. With this aim the rigorous examination of the frequency response and the antenna sensitivity pattern for the massless scalar waves was performed in \cite{Nakao01} in the whole frequency domain; in \cite{MagNic00} the detector response was analyzed in the conformal and in the synchronous gauge in the long wavelength limit and the equivalence of the two approaches was demonstrated.

In the paper we studied this basic issues in the case of the massive scalar wave. This kind of perturbations can arise in a number of alternative theories, in particular they can be realized in the massive Branse--Dicke theory or in
a class of the $f(R)$ extended theories of gravity \cite{CCL08}, \cite{LaurCap2011}. The detector response for this case was studied in \cite{MagNic00} in the long wavelength limit and in \cite{CCL08} for the full frequency spectrum.
Here we carry out the analysis of the response of the laser interferometer to the scalar wave in the conformal gauge and in the synchronous gauge for all frequencies and we show the equivalence of the two approaches. We show as well the equivalence of this two gauges on the level of solutions of the linearized field equations of the massive Brans--Dicke theory. We present basic angular and frequency characteristics of the gravitational wave antenna. The response of the detector written in the synchronous gauge is particularly useful since in this coordinates the free motion of test bodies (like beam splitter, mirrors etc.) is simple. Thus although the response was explicitly given for the static interferometer the motion of the detector can straightforwardly be taken into account. This analysis can be applied to currently working Earth--based detectors but in particularly to the future, next generation experiments like Einstein Telescope or space--based missions like eLISA where it may be also essential to go beyond the long--wavelength approximation of the interferometer response which is usually assumed when working in the local Lorentz frame.

\section{Acknowledgments}

The work was supported in part by the National Science Centre grants UMO-2013/01/ASPERA/ST9/00001
and UMO-2014/14/M/ST9/00707.

\appendix

\section{Scalar polarization modes}

\label{A1}

Let the orthonormal basis $\{{\bf e}_{x},{\bf e}_{y},{\bf e}_{z}\equiv\bO\}$ represents the source frame and $\bf n$ be the unit vector
along the detector arm, then
\begin{eqnarray}
\label{eq:polA1}
\be^{sl} & = & {\bf e}_{z}\otimes{\bf e}_{z} = \bO\otimes\bO \\
\be^{st} & = & {\bf e}_{x}\otimes{\bf e}_{x} + {\bf e}_{y}\otimes{\bf e}_{y} \nonumber
\end{eqnarray}
thus
\begin{eqnarray}
\label{eq:polA2}
{\bf n}\otimes{\bf n}:\be^{sl} & = & \left({\bf n}\cdot\bO\right)^2 \\
{\bf n}\otimes{\bf n}:\be^{st} & = & \left({\bf n}\cdot{\bf e}_{x}\right)^2 + \left({\bf n}\cdot{\bf e}_{y}\right)^2 =
1 - \left({\bf n}\cdot\bO\right)^2. \nonumber
\end{eqnarray}
We can now identify an arbitrary Cartesian coordinates having $\{{\bf e}_{x},{\bf e}_{y},{\bf e}_{z}\equiv\bO\}$ as an orthonormal basis
with the spatial part of our synchronous coordinate system. We can aslo relate vectors and covectors in the canonical way:
$\Omega_i=\Omega^j\delta_{ij}=\Omega^i$. Therefore in an arbitrary synchronous coordinates the spatial tensors $\be^{st}$, $\be^{sl}$
can be written as
\begin{eqnarray}
\label{eq:polA3}
\be^{sl} & = & \Omega_i\,\Omega_j\;dx^i dx^j \\
\be^{st} & = & \left(\delta_{ij} - \Omega_i\,\Omega_j\right)dx^i dx^j . \nonumber
\end{eqnarray}

\section{Gravitational waves in the massive Brans-Dicke theory in the synchronous gauge}

\label{A2}

In this Appendix we obtain the solution of the linearized field equations (\ref{eq:ein:lin}), (\ref{eq:BD:lin})
in the gauge $h_{\mu0}=0$. To this end we recall the explicit form of $R_{\mu\alpha\nu\beta}^{(1)}$ \cite{MTW}:
\begin{eqnarray}
\label{eq:riem}
R^{(1)}_{\mu\alpha\nu\beta} & = & \frac12 \left[ h_{\mu\beta,\nu\alpha} + h_{\nu\alpha,\mu\beta} -
h_{\mu\nu,\alpha\beta} - h_{\alpha\beta,\mu\nu} \right].
\end{eqnarray}
Now we chose the gauge $h_{\mu0}=0$ and we assume the plane wave solutions $h_{ij}(t,z)$, $\Phi(t,z)$ for a wave propagating
along the $z$ direction; taking the spatial trace of the field Eqs. (\ref{eq:ein:lin}) and using Eq. (\ref{eq:BD:lin}) we obtain
\begin{eqnarray}
\label{eq:h123}
- \Box_2 h_{11} - \Box_2 h_{22} + h_{33,tt} + 3\,m^2\Phi = 0,\qquad\text{where }\quad
\Box_2\equiv -\partial_{t}^2 + \partial_{z}^2.
\end{eqnarray}
Substituting $h_{33,tt}$ obtained in Eq. (\ref{eq:h123}) to '$11$' and '$22$' components of Eqs. (\ref{eq:ein:lin}) one gets
\begin{eqnarray}
\label{eq:h11}
\Box_2 h_{11} & = & m^2\Phi \\
\label{eq:h22}
\Box_2 h_{22} & = & m^2\Phi,
\end{eqnarray}
with the solutions $h_{11} = \Phi + h_{11}^0$, $h_{22} = \Phi + h_{22}^0$, where
$ h_{11}^0$ and  $h_{22}^0$ solve the homogeneous wave equation; in turn '$00$' component
of (\ref{eq:ein:lin}),
\begin{eqnarray}
\label{eq:h00}
- \frac12 h_{11,zz} - \frac12 h_{22,zz} + m^2\Phi + \Phi_{tt} = 0,
\end{eqnarray}
impose the trace--free condition for the homogeneous solutions, $h_{+}\equiv h_{11}^{0} = -h_{22}^{0}$.
From '$33$' part of the system (\ref{eq:ein:lin}) using Eqs. (\ref{eq:h123}), (\ref{eq:h22}) one obtains
\begin{eqnarray}
\label{eq:h33}
h_{33,tt} = - m^2\Phi,
\end{eqnarray}
whereas '$12$' part gives
\begin{eqnarray}
\Box_2 h_{12}=0\;;
\end{eqnarray}
the remaining equations (\ref{eq:ein:lin}) are identities or show that $h_{13}=h_{23}=0$.
Thus the full set of modes consists of two standard massless helicity $2$ states $h_{+}$, $h_{\times}\equiv h_{12}$
and two massive modes (but one degree of freedom), $h_{st}=\Phi$, $h_{sl}\equiv h_{33}$.

\section{Equivalence of the conformal and synchronous gauges}

\label{A3}

In this Appendix we show the equivalence of the two gauges, the conformal gauge, Eq. (\ref{eq:conf}) and the synchronous gauge,
Eq. (\ref{eq:synch}). Note that under $3$--dimensional rigid rotations in the conformal coordinates components of the gravitational
wave tensor $h_{s}(t,{\bf x})\eta_{\mu\nu}$ transforms as a set of four scalar fields whereas in the synchronous gauge
the diagonal form of the tensor is not preserved thus one expects off-diagonal terms of the scalar modes as well which will mix with the
transveres traceless modes. To simplify the calculations we will treat here only the scalar modes;
we will work in the frame in which the wave propagates in an arbitrary direction $\bO$;

First we assume the gravitational wave of the form $h_{s}(t-\bO\cdot{\bf x}/{\rm v})$. In this case the following transformation of coordinates
\begin{eqnarray}
\label{eq:conf2synch1}
{\bf x} & = & {\bf x}' + \frac12{\bf f}(t'-\bO\cdot{\bf x}'/{\rm v}),\qquad
{\bf f}(t'-\bO\cdot{\bf x}'/{\rm v}) = \frac{\bO}{{\rm v}}\int\limits_{-\infty}^{t'-\frac{\bO\cdot{\bf x}'}{{\rm v}}}h_{s}(v)dv,\\
t & = & t' + \frac12 g(t'-\bO\cdot{\bf x}'/{\rm v}),\qquad g(t'-\bO\cdot{\bf x}'/{\rm v}) =
- \int\limits_{-\infty}^{t'-\frac{\bO\cdot{\bf x}'}{{\rm v}}}h_{s}(v)dv, \nonumber
\end{eqnarray}
gives
\begin{eqnarray}
\label{eq:conf2synch2}
d{\bf x} & = & d{\bf x}' + \frac{\bO}{2{\rm v}}h_{s}(t'-\bO\cdot{\bf x}'/{\rm v})\,dt'
- \frac{\bO}{2{\rm v}^2}h_{s}(t'-\bO\cdot{\bf x}'/{\rm v})\,\left(\bO\cdot d{\bf x}'\right)\\
dt & = & dt' - \frac{1}{2}h_{s}(t'-\bO\cdot{\bf x}'/{\rm v})\,dt'
+ \frac{1}{2{\rm v}}h_{s}(t'-\bO\cdot{\bf x}'/{\rm v})\,\left(\bO\cdot d{\bf x}'\right) \nonumber
\end{eqnarray}
which leads to
\begin{eqnarray}
\label{eq:conf2synch3}
\left(1+h_{s}\right)\eta_{\mu\nu}dx^{\mu}dx^{\nu} & = & \eta_{\mu\nu}dx'^{\mu}dx'^{\nu} + h_{s}\left(dx'^2 + dy'^2\right)
-\frac{\Omega_i\,\Omega_j}{{\rm v}^2}h_{s}\,dx'^idx'^j \\
& = & \eta_{\mu\nu}dx'^{\mu}dx'^{\nu} + \be^{st} h_{s} + \left(1-\frac{1}{{\rm v}^2}\right)\be^{sl}h_{s}, \nonumber
\end{eqnarray}
according to Eq. (\ref{eq:polA3}).

In the general case of a superposition of monochromatic waves $e^{i\omega(t-\bO\cdot{\bf x}/{\rm v})}$,
$$
{\rm h}_{s}(t,{\bf x})=\int\limits_{m}^{\infty}\,d\omega
\int\limits_{S^2}d\bO \, e^{i\omega\left[t-\bO\cdot{\bf x}/{\rm v}(\omega,\bO)\right]} \chi(\omega,\bO),
$$
where the form of a possibly orientation--dependent dispersion relation ${\rm v}(\omega,\bO)$ is dictated by
the field equations for ${\rm h}_{s}$ the generators ${\bf f}$, $g$ of the gauge transformation, Eq. (\ref{eq:conf2synch1}), 
are given respectively by
\begin{eqnarray}
\label{eq:fg}
\frac{\bO}{2{\rm v}}\int\limits_{m}^{\infty}\,d\omega
\int\limits_{S^2}d\bO \; {\bf f}\left[t'- \bO\cdot{\bf x}'/{\rm v}(\omega,\bO)\right],\qquad
-\int\limits_{m}^{\infty}\,d\omega
\int\limits_{S^2}d\bO \; g\left[t'- \bO\cdot{\bf x}'/{\rm v}(\omega,\bO)\right]. \nonumber
\end{eqnarray}
and then the scalar modes in the synchronous gauge read
\begin{eqnarray}
\int\limits_{m}^{\infty}\,d\omega
\int\limits_{S^2}d\bO \,\left[ \be^{st}(\bO) + \left(1-\frac{1}{{\rm v^2(\omega,\bO)}}\right)
\be^{sl}(\bO) \right] e^{i\omega\left[t-\bO\cdot{\bf x}/{\rm v(\omega,\bO)}\right]}\chi(\omega,\bO).
\end{eqnarray}

As an example let us consider the massive Brans--Dicke theory; in the conformal coordinates we have
\begin{eqnarray}
\label{eq:BD1}
\Box h_{s} = m^2 h_{s}, \qquad
{\rm v}(\omega) = \frac{|\omega|}{\sqrt{\omega^2 - m^2}},\qquad 1-\frac{1}{{\rm v}^2(\omega)} & = & \frac{m^2}{\omega^2} \nonumber.
\end{eqnarray}
The gauge transformation (\ref{eq:conf2synch1}) for the plane wave propagating in the $-z$ direction,
$h_{s}=\int\limits_{-\infty}^{\infty}\;d\omega\; e^{i\omega(t+z/{\rm v})}\chi(\omega)$, with
\begin{eqnarray}
\label{eq:fg3}
{\bf f} = \left(0,0,\int\limits_{m}^{\infty}\,d\omega \frac{i}{\omega\,{\rm v}} e^{i\omega(t+z/{\rm v})}\chi(\omega)\right),
\qquad g = \int\limits_{m}^{\infty}\,d\omega \frac{i}{\omega} e^{i\omega(t+z/{\rm v})}\chi(\omega)
\end{eqnarray}
connects then
\begin{eqnarray}
\label{eq:synchA1}
{\rm \bf h}(t,z) & = &
\text{diag}\left( - {\rm h}_s(t,z), {\rm h}_s(t,z), {\rm h}_s(t,z), {\rm h}_s(t,z) \right)
\end{eqnarray}
with
\begin{eqnarray}
\label{eq:synchA2}
{\rm \bf h}'(t',z') & = &
\text{diag}\left(0, {\rm h}_s(t',z'), {\rm h}_s(t',z'), \frac{m^2}{-\partial_{z'}^2 + m^2}{\rm h}_{s}(t',z')\right).
\end{eqnarray}
We see that the scalar mode has the same form as the solutions to the linearized field equations of the massive Brans--Dicke model
obtained in the Appendix \ref{A2}. Note that for $m=0$ the generators of the gauge transformation given in Eqs. (\ref{eq:conf2synch1})
or (\ref{eq:fg3}) reduce to the transformation given in \cite{MagNic00}, Eqs. (B$5$, B$6$).
We stress however that the result given here is more general: for any mode $h_{s}$ that in the conformal gauge is constrained by the
field equation $\partial_{t}^2h_{s} = F[\partial_{z}]h_{s}$ the corresponding synchronous--gauge form of the wave (in the  frame where the wave propagates along the $z$ axis) reads
\begin{eqnarray}
\label{eq:synchA3}
{\rm \bf h}(t,z) & = &
\text{diag}\left(0, {\rm h}_s(t,z), {\rm h}_s(t,z),
\frac{F\left[\partial_{z}\right] - \partial_{z}^2}{F\left[\partial_{z}\right]}{\rm h}_{s}(t,z)\right).
\end{eqnarray}



\end{document}